\documentclass[aps,pra,reprint,superscriptaddress,nofootinbib,notitlepage,amsmath,amssymb,showpacs]{revtex4-1}

\usepackage{graphicx}
\usepackage{dcolumn}
\usepackage{bm}
\usepackage[colorlinks,citecolor=blue]{hyperref}
\usepackage[figure,figure*]{hypcap}
\usepackage{romannum}

\begin{document}

\title{Trapping Ultracold Atoms in a Sub-Micron-Period Triangular Magnetic Lattice}

\author{Y. Wang}
\affiliation{Centre for Quantum and Optical Science, Swinburne University of Technology, Hawthorn, Victoria 3122, Australia}

\author{T. Tran}
\affiliation{Centre for Quantum and Optical Science, Swinburne University of Technology, Hawthorn, Victoria 3122, Australia}

\author{P. Surendran}
\affiliation{Centre for Quantum and Optical Science, Swinburne University of Technology, Hawthorn, Victoria 3122, Australia}

\author{I. Herrera}
\affiliation{Centre for Quantum and Optical Science, Swinburne University of Technology, Hawthorn, Victoria 3122, Australia}
\affiliation{Department of Physics, University of Auckland, Private Bag 92019, Auckland, New Zealand}

\author{A. Balcytis}
\affiliation{Centre for Micro-Photonics, Swinburne University of Technology, Hawthorn, Victoria 3122, Australia}
\affiliation{Melbourne Centre for Nanofabrication, Victorian Node of the Australian National Fabrication Facility, 151 Wellington Rd., Clayton, Victoria 3168, Australia}
\affiliation{Centre for Physical Sciences and Technology, Savanoriu Ave 2131, LT-02300 Vilnius, Lithuania}

\author{D. Nissen}
\affiliation{Experimental Physics IV, Institute of Physics, University of Augsburg, Universit\"{a}tstrasse 1, D-86159 Augsburg, Germany}

\author{M. Albrecht}
\affiliation{Experimental Physics IV, Institute of Physics, University of Augsburg, Universit\"{a}tstrasse 1, D-86159 Augsburg, Germany}

\author{A. Sidorov}
\affiliation{Centre for Quantum and Optical Science, Swinburne University of Technology, Hawthorn, Victoria 3122, Australia}

\author{P. Hannaford}
\affiliation{Centre for Quantum and Optical Science, Swinburne University of Technology, Hawthorn, Victoria 3122, Australia}

\date{\today}

\begin{abstract}
We report the trapping of ultracold $^{87}$Rb atoms in a $0.7\,\mu$m-period two-dimensional triangular magnetic lattice on an atom chip. The magnetic lattice is created by a lithographically patterned magnetic Co/Pd multilayer film plus bias fields. Rubidium atoms in the $|F=1,m_{F}=-1\rangle$ low-field seeking state are trapped at estimated distances down to about $100\,$nm from the chip surface and with calculated mean trapping frequencies up to about $800\,$kHz. The measured lifetimes of the atoms trapped in the magnetic lattice are in the range 0.4 - $1.7\,$ms, depending on distance from the chip surface. Model calculations suggest the trap lifetimes are currently limited mainly by losses due to one-dimensional thermal evaporation following loading of the atoms from the Z-wire trap into the very tight magnetic lattice traps, rather than by fundamental loss processes such as surface interactions, three-body recombination or spin flips due to Johnson magnetic noise. The trapping of atoms in a $0.7\,\mu$m-period magnetic lattice represents a significant step towards using magnetic lattices for quantum tunneling experiments and to simulate condensed matter and many-body phenomena in nontrivial lattice geometries.

\end{abstract}


\maketitle

\section{INTRODUCTION}
Magnetic lattices consisting of periodic arrays of microtraps created by patterned magnetic films on an atom chip provide a potential complementary tool to optical lattices for simulating condensed matter and many-body phenomena (e.g.,~\cite{Yibo2016}). Such lattices have a high degree of flexibility and may, in principle, be fabricated with almost arbitrary two-dimensional (2D) and one-dimensional (1D) geometries and lattice spacing~\cite{Schmied2010} and may be readily scaled up. In addition, magnetic lattices do not require high power, stable laser beams and precise beam alignment, they operate with relatively little technical noise, power consumption, or heating, and they involve state-selective atom trapping, allowing rf evaporative cooling to be performed in the lattice and rf spectroscopy to be used to characterize the lattice-trapped atoms \textit{in situ}. Finally, magnetic lattices have the potential to enable miniaturized integrated quantum technologies exploiting many-body states of ultracold atoms and hybrid quantum systems such as quantum registers with on-chip readout.

However, magnetic lattices are still in their infancy compared with optical lattices due largely to the difficulty in fabricating high-quality magnetic microstructures, especially lattices with sufficiently small periods to enable quantum tunneling experiments. To date, 1D magnetic lattices~\cite{Singh2008,Jose2014,Surendran2015} and 2D rectangular~\cite{Gerritsma2007,Whitlock2009}, square~\cite{Leung2011,Herrera2016} and triangular~\cite{Leung2011,Herrera2016} magnetic lattices with periods down to $10\,\mu$m have been produced and clouds of ultracold atoms have been trapped in them~\cite{Singh2008,Jose2014,Surendran2015,Gerritsma2007,Whitlock2009,Leung2014}. In the case of the $10\,\mu$m-period 1D magnetic lattice, $^{87}$Rb atoms have been cooled to degeneracy to create a periodic array of isolated Bose-Einstein condensates~\cite{Jose2014,Surendran2015}. In order to conduct experiments involving quantum tunneling, lattices with periods in the sub-micron regime are required (e.g.,~\cite{Bloch2008,Bakr2009}).

In this paper we report the trapping of ultracold $^{87}$Rb $|F=1,m_{F}=-1\rangle$ atoms in a $0.7\,\mu$m-period triangular magnetic lattice on an atom chip. The magnetic lattice is created by a lithographically patterned magnetic Co/Pd multilayer film plus bias fields~\cite{Herrera2016}. The design of the triangular magnetic lattice and calculations of the lattice trapping potentials including the effect of the Casimir-Polder surface interaction are presented in Sec.~\ref{2}. Sec.~\ref{3} gives experimental details, including the fabrication and characterization of the $0.7\,\mu$m-period triangular magnetic lattice structure. In Sec.~\ref{4} we present experimental results for the interaction of the ultracold atoms with the magnetic lattice potential, loading of atoms into the magnetic lattice traps, and lifetime measurements of the lattice-trapped atoms at various distances from the chip surface. In Sec.~\ref{5} we discuss possible ways for improving the lifetimes and the loading procedure, and in Sec.~\ref{6} we summarize our results.
\begin{figure*}
\includegraphics[width=460pt]{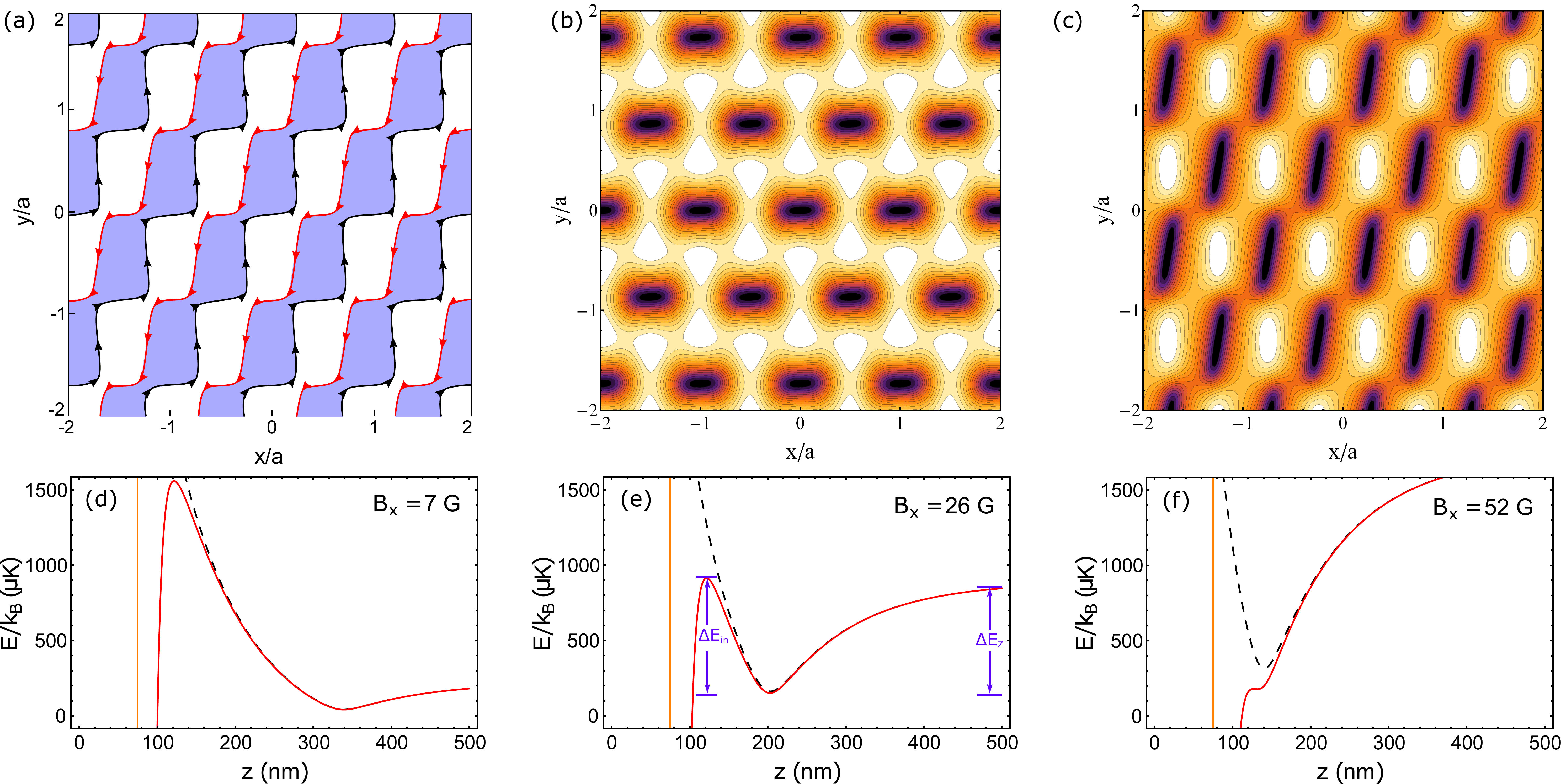}
\caption{(a) Magnetic film pattern designed to create a triangular magnetic lattice optimised for a trap distance $z = z_{min} = a/2$ from the surface of the magnetic film, where $a$ is the lattice period.  Blue regions represent the magnetic film and arrows represent virtual currents circulating around the edges of the film structure. (b) Contour plot of the optimized triangular magnetic lattice potential with bias fields $B_{x} = 0.5\,$G, $B_{y} = 4.5\,$G; $a = 0.7\,\mu$m; and $z_{min} = a/2 = 350\,$nm. Dark regions are trap minima. (c) Contour plot of a triangular magnetic lattice potential with bias fields $B_{x} = 52\,$G, $B_{y} = 0\,$G; $a = 0.7\,\mu$m; and $z_{min} = 139\,$nm.
(d-f) Calculated trapping potentials for $^{87}$Rb $|F=1, m_{F} = -1\rangle$ atoms trapped in a $0.7\,\mu$m-period triangular magnetic lattice for bias fields $B_{x}$ = (d) $7\,$G; (e) $26\,$G; (f) $52\,$G. Black dashed lines are the magnetic lattice potentials and red solid lines include the Casimir-Polder interaction with $C_{4} = 8.2 \times 10^{-56}\,$Jm$^{4}$ for a silica surface. Vertical orange lines indicate the position of the silica surface ($z = 75\,$nm) used in the calculations. Magnetic film parameters: magnetization $4\pi M_{z} = 5.9\,$kG, film thickness $t_{m} = 10.3\,$nm.
}
\label{Fig1}
\end{figure*}

\section{THE SUB-MICRON-PERIOD TRIANGULAR MAGNETIC LATTICE}
\label{2}
The triangular magnetic lattice structure is designed using the linear programming algorithm developed by Schmied et al.~\cite{Schmied2010}. Figure~\ref{Fig1}(a) shows the magnetic film pattern designed to create a triangular lattice optimized for a trap distance $z = z_{min} = a/2$ from the surface of the magnetic film, where $a$ is the lattice period. For $a  = 0.7\,\mu$m and a film with perpendicular magnetization $4\pi M_{z} = 5.9\,$kG (or $M_{z} = 470\,$emu/cm$^{3}$) and nominal thickness $t_{m} = 10.3\,$nm, the required bias magnetic fields are $B_{x} = 0.5\,$G, $B_{y} = 4.5\,$G, where the $x$- and $y$-directions are defined in Fig.~\ref{Fig1}. A 2D contour plot for these parameters is shown in Fig.~\ref{Fig1}(b). In the present experiment, the magnetic lattice is loaded with atoms from a Z-wire magnetic trap operating with a bias field $B_{x} \approx 52\,$G (parallel to the ends of the Z-wire). Figure~\ref{Fig1}(c) shows a 2D contour plot for the $0.7\,\mu$m-period triangular lattice structure with bias fields $B_{x} = 52\,$G, $B_{y} = 0$ and the above parameters. For this magnetic lattice, the traps are more elongated and tighter than for the optimized triangular lattice with $B_{x} = 0.5\,$G, $B_{y} = 4.5\,$G and each trap is surrounded by four rather than six potential maxima.

For a magnetic film structure magnetized in the $z$-direction, the magnetization can be modeled as a virtual current circulating around the edges of the patterned structure, as indicated by the arrows in Fig.~\ref{Fig1}(a). A bias field $B_{y}$ applied along the +$y$-direction can cancel the magnetic field produced by the virtual current flowing along the horizontal black edge of the patterned structure shown in Fig.~\ref{Fig1}(a) to create a periodic array of magnetic traps aligned along the short horizontal black edges (Fig.~\ref{Fig1}(b)). On the other hand, a bias field $B_{x}$ applied along the +$x$-direction can cancel the magnetic field produced by the virtual current flowing along the vertical red edge to create a periodic array of elongated magnetic traps aligned along the long vertical red edges (Fig.~\ref{Fig1}(c)). In general, a larger bias field $B_{x}$ produces lattice traps which are closer to the magnetic film, and which are tighter and deeper.

\begin{table*}
\caption{Calculated parameters for $^{87}$Rb $|F=1, m_{F}=-1\rangle$ atoms trapped in the $0.7\,\mu$m-period triangular magnetic lattice, for $4\pi M_{z} = 5.9\,$kG, $t_{m} = 10.3\,$nm, $C_{4} = 8.2 \times 10^{-56}\,$Jm$^{4}$, and offset parameter $\delta d = 25\,$nm (see Sec.~\ref{4}D). $z_{min}$ and $d = (z_{min} - 50)\,$nm are the distances of the trap minima from the magnetic film surface and the chip surface, respectively; $B_{IP}$ is the trap bottom; $\omega_{\perp}$, $\omega_{\parallel}$ are the trap frequencies perpendicular to and parallel to the elongated traps; $\overline{\omega}$ is the geometric mean trap frequency; $\Delta E_{x,y}$, $\Delta E_{z}$ are the barrier heights of the magnetic potential in the $x$-$y$ plane and $z$-direction, respectively; and $\Delta E_{in}$ is the barrier height including the effect of the Casimir-Polder interaction.}
\vspace{0.2cm}
\centering
{\begin{tabular}{|c|c|c|c|c|c|c|c|c|}
\hline
Bias field  &  $z_{min}$ & $d$ & $B_{IP}$ & $\omega_{\perp, \parallel}/2\pi$ & $\overline{\omega}/2\pi$ & $\Delta E_{x,y}/k_{B}$ & $\Delta E_{z}/k_{B}$ & $\Delta E_{in}/k_{B}$ \\
$B_{x}$ (G) & (nm) & (nm) & (G) & (kHz) & (kHz) & ($\mu$K) & ($\mu$K) & ($\mu$K)\\ \cline{1-9}
7 & 339 & 289 & 1.2 & 532, 82 & 285 & 264, 170 & 181 & 2348 \\ \hline
9 & 310 & 260 & 1.6 & 618, 94 & 330 & 359, 232 & 244 & 2258 \\ \hline
14 & 267 & 217 & 2.5 & 772, 115 & 409 & 559, 362 & 376 & 2072 \\ \hline
26 & 203 & 153 & 4.7 & 1097, 153 & 569 & 1104, 729 & 731 & 1584 \\ \hline
40 & 163 & 113 & 6.7 & 1405, 185 & 715 & 1703, 1155 & 1118 & 1075 \\ \hline
52 & 139 & 89 & 8.2 & 1657, 207 & 828 & 2233, 1554 & 1465 & 655 \\ \hline
\end{tabular}}
\label{T1}
\end{table*}

For the $0.7\,\mu$m-period magnetic lattice, the atoms are trapped at distances down to about $100\,$nm from the chip surface, so that effects of surface interactions need to be considered. The trapping potential at distance $z$ from the magnetic film surface may be expressed as
\begin{equation}
\label{CP}
V(z) = V_{M}(z) + V_{CP}(d),
\end{equation}
where $V_{M}(z)$ is the magnetic lattice potential, $V_{CP}(d)$ is the combined Casimir-Polder and van der Waals potential, and $d = z_{min}-(t_{Au}+t_{SiO_{2}})$ is the distance of the trap centre from the surface of the atom chip (allowing a thickness $(t_{Au}+t_{SiO_{2}}) = 75\,$nm for the gold and silica surface layers in the present experiment). $V_{CP}(d)$ may be expressed as (e.g.,~\cite{Pasquini2004})
\begin{equation}
\label{CP1}
V_{CP}(d) = - \frac{C_{4}}{d^{3}(d+3\lambda_{opt}/2\pi^{2})},
\end{equation}
where $C_{4}=\frac{1}{4\pi \epsilon_{0}}\frac{3\hbar c \alpha_{0}}{8\pi}\frac{\epsilon_{r}-1}{\epsilon_{r}+1}\phi(\epsilon_{r})$~\cite{Lin2004} is the Casimir-Polder coefficient, $\alpha_{0}$ is the static atomic polarizability, $\phi(\epsilon_{r})$ is a numerical factor~\cite{Yan1997} that depends on the relative permittivity $\epsilon_{r}$ of the top surface layer, $\epsilon_{0}$ is the vacuum permittivity, and $\lambda_{opt}$ is the wavelength of the strongest electric dipole transition of the atom. The gravitational potential is negligible compared with the strong magnetic lattice potential and is not included in Eq.~(\ref{CP}).

Figures~\ref{Fig1}(d)-(f) present calculations of the trapping potentials for different bias fields $B_{x}$, where $C_{4}$ is taken to be $8.2\times10^{-56}\,$Jm$^{4}$ for a dielectric surface of silica film, for which $\epsilon_{r}= 4.0$ and $\phi(\epsilon_{r}) = 0.771$, and $\alpha_{0}  = 5.25 \times 10^{-39}\,$Fm$^{2}$ for a ground-state Rb atom. The vertical orange lines in Fig.~\ref{Fig1} (d)-(f) indicate the position of the chip surface which is taken here to be $75\,$nm from the magnetic film. According to these calculations, the trapping potential for $B_{x} = 52\,$G is very shallow (trap depth $\Delta E_{in}/k_{B} \sim 1.5\,\mu$K). Introducing an offset $\delta d = + 25\,$nm (see Sec.~\ref{4}D) for the distance $d = z_{min}-(t_{Au}+t_{SiO_{2}})$ of the lattice traps from the chip surface gives $\Delta E_{in}/k_{B} = 655\,\mu$K for $B_{x} = 52\,$G.  The calculated trap parameters for different bias fields $B_{x}$ with $\delta d = 25\,$nm are listed in Table~\ref{T1}. For $B_{x}<26\,$G, the trap centre is located at distances $d > 150\,$nm from the chip surface and the effect of the Casimir-Polder interaction is small, so that the effective depth of the lattice traps $\Delta E_{eff} \equiv \Delta E_{z}$. For $B_{x}>40\,$G, the trap centre is located $<110\,$nm from the chip surface, depending on the distance $d$, and the magnetic potential is deformed by the attractive Casimir-Polder interaction, so that $\Delta E_{eff} \equiv \Delta E_{in}$. For these very tight magnetic lattice traps, the atom densities are very high; for example, for $B_{x} = 26\,$G and assuming two atoms per lattice site, the calculated peak atom density is $n_{0} \approx 2 \times 10^{15}\,$cm$^{-3}$.

\section{EXPERIMENT}
\label{3}
\subsection{Fabrication of the 0.7 \texorpdfstring{$\mu$m}{Lg}-period triangular magnetic lattice structure}
The magnetic film used for fabrication of the $0.7\,\mu$m-period magnetic lattice structure consists of a stack of eight bi-layers of alternating Pd ($0.9\,$nm) and Co ($0.28\,$nm)~\cite{Herrera2016,Stark2015}. Such multilayer films have a large perpendicular magnetic anisotropy and a high degree of magnetic homogeneity is expected. In addition, they exhibit a large saturation magnetization ($4\pi M_{z} = 5.9\,$kG), square-shaped hysteresis loops~\cite{Herrera2016}, a high coercivity ($H_{c} \sim 1\,$kOe), a high Curie temperature (300 - $400\,^{\circ}$C) and a very small grain size (down to $\sim6\,$nm). Alternating layers of $0.9\,$nm Pd and $0.28\,$nm Co are known to exhibit an enhanced ($\sim$ 20$\%$) magnetization relative to bulk cobalt due to polarization of the Pd atoms by the nearby Co layers (e.g.,~\cite{Stinson1990}).

The Co/Pd multilayers are deposited by dc-magnetron sputtering onto a seed layer of $3\,$nm-thick Pd plus $3\,$nm-thick Ta on a $500\,\mu$m-thick Si(100) substrate~\cite{Herrera2016}. A $1.1\,$nm protective layer of Pd is deposited on top of the Co/Pd stack. The active magnetic thickness of the stack is taken to be $t_{m}= 10.3\,$nm\footnote{In~\cite{Herrera2016}, the active magnetic thickness of the Co/Pd stack was given as $2.24\,$nm, which represents the total Co thickness only.}, where an additional $0.9\,$nm Pd is included to allow for polarization of the $3\,$nm Pd in contact with the bottom Co layer.

The $0.7\,\mu$m-period triangular magnetic lattice structure was fabricated using electron-beam lithography (EBL) plus reactive ion etching~\cite{Herrera2016}. A $300\,$nm-thick layer of positive tone resist (PMMA 495k polymer, MicroChem Corp) is spin-coated onto a Co/Pd film-coated silicon wafer and the triangular lattice pattern (Fig.~\ref{Fig1}(a)) is written onto the resist using an e-beam lithography machine operating at $100\,$kV (Raith EBPG5000plusES). A $5\,$nm electron spot is scanned along the designated pattern at $50\,$MHz rate by the pattern generator. The $1\,$mm$^{2}$ write field of the e-beam machine allows exposure of an entire magnetic lattice structure without the need to move the sample stage. When a uniform EBL exposure is performed over a large ($1\,$mm$^{2}$) area, the electron beam can be scattered in the resist to produce a pattern that is deformed towards the edges. To compensate for these proximity effects an exposure dose proximity map is designed using Monte-Carlo simulations to evaluate the scattering of the electron beam~\cite{Yibo2017}. The duration of the EBL exposure is about two hours. After development of the resist, the triangular pattern is etched into the Co/Pd film by argon-ion bombardment in an inductively-coupled plasma reactive ion etching tool (Samco RIE-101iPH).

\begin{figure}
\begin{centering}
\includegraphics[width=245pt]{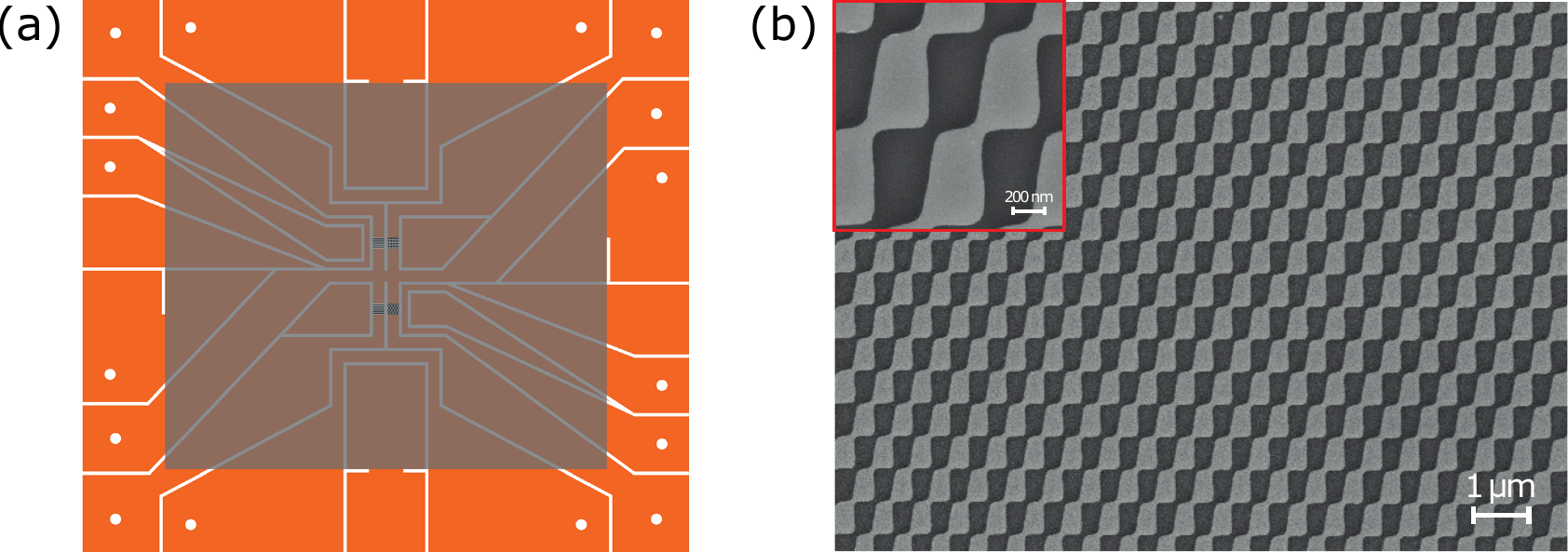}
\caption{(a) Schematic of the DBC atom chip. The structure includes four separated current-carrying U-wire and Z-wire structures for trapping the ultracold atom cloud and loading into the magnetic lattice traps plus two wires on either side for rf evaporative cooling or rf spectroscopy. The small green squares in the center show the positions of four magnetic lattice structures, which are located below their respective U- and Z-wires. (b) SEM image of the fabricated $0.7\,\mu$m-period triangular magnetic lattice structure. Light-gray regions are the unetched magnetic film and the dark-gray regions are the etched regions.}
\par\end{centering}
\label{Fig2}
\end{figure}

\begin{table}
\caption{Nominal values of magnetic lattice parameters.}
\vspace{0.2cm}
\begin{centering}
\begin{tabular}{|c|c|c|}
\hline
Parameter & Symbol & Nominal value \\ \hline
Lattice period & $a$ & 0.70 $\mu$m \\ \hline
Remanent magnetization & $4\pi M_{z}$ & $5.9\,$kG \\ \hline
Active magnetic film thickness & $t_{m}$ & $10.3\,$nm \\ \hline
Thickness of Au surface layer & $t_{Au}$ & $50\,$nm \\ \hline
Thickness of SiO$_{2}$ surface layer & $t_{SiO_{2}}$ & $25\,$nm \\ \hline
\end{tabular}
\par\end{centering}
\label{T2}
\end{table}
The patterned Co/Pd magnetic film is coated with a reflective $50\,$nm layer of gold plus a $25\,$nm layer of silica to prevent rubidium atoms reacting with the gold surface. The patterned Co/Pd magnetic film is then glued onto a direct bonded copper (DBC) $50\,\mathrm{mm}\times55\,$mm atom chip~\cite{Squires2011} comprising $130\,\mu$m-thick current-carrying U-wire and Z-wire structures~\cite{Yibo2017}. The atom chip can accommodate four separate $1\,$mm$^{2}$ magnetic lattice structures, each of which has a U-wire and Z-wire structure directly beneath it (Fig.~\ref{Fig2}(a)).

Finally, the $0.7\,\mu$m-period Co/Pd triangular magnetic lattice structure is magnetized and then characterized by magnetic/atomic force microscopy and scanning electron microscopy, prior to mounting in the vacuum chamber. The period of the triangular magnetic structure is measured from scanning electron microscopy (SEM) scans (Fig.~\ref{Fig2}(b)) to be $0.70\,\mu$m within about 1$\%$. The quality of the present $0.7\,\mu$m-period triangular magnetic lattice structure is significantly improved over that reported earlier~\cite{Herrera2016}.

The parameters of the triangular magnetic lattice structure are summarized in Table~\ref{T2}.

\subsection{Atom trapping and cooling and atom imaging}
Rubidium atoms released from a pulsed dispenser are trapped in a standard four-beam mirror magneto-optical trap (MMOT) on the atom chip with a gold reflecting surface. The beams derived from a $1\,$W tapered amplifier laser system consist of an atom trapping beam detuned $15\,$MHz below the $F = 2 \rightarrow F^{\prime} = 3$ cycling transition combined with a repumper beam locked to the $F = 1 \rightarrow F^{\prime} = 2$ transition. We trap typically $2 \times 10^{8}$ atoms in $25\,$s in the MMOT at 1 - $2\,$mm below the chip surface. The atoms are then transferred to a compressed MMOT formed by passing $20\,$A through a U-wire on the atom chip plus a bias field $B_{x} = 12\,$G to create the quadrupole trap. This is followed by a polarization gradient cooling stage, resulting in $\sim 1.5 \times 10^{8}$ atoms cooled to $\sim 40\,\mu$K.

The atoms are then optically pumped to the required $|F=1, m_{F}=-1\rangle$ low-field seeking ground state, which is chosen because of its smaller three-body recombination rate~\cite{Burt1997,Soding1999} compared with the $|F=2, m_{F}=+2\rangle$ state. Next, the atoms are transferred to a Z-wire magnetic trap formed by passing a current $I_{z} = 35\,$A and raising the bias field to $B_{x} = 33\,$G. The trap bottom is adjusted to $\sim3\,$G to prevent spin-flip loss by applying a bias field $B_{y} = 7\,$G. To enhance the elastic collision rate, the atom cloud is then compressed by ramping $I_{z}$, $B_{x}$ and $B_{y}$ up to $37\,$A, $52\,$G and $8\,$G, respectively, in $100\,$ms resulting in $\sim 5 \times 10^{7}$ atoms at a temperature of $\sim 200\,\mu$K at $\sim 700\,\mu$m below the chip surface with a Z-wire trap lifetime of $\sim 20\,$s. Forced rf evaporative cooling is then applied to the atoms in the Z-wire trap for $12\,$s by logarithmically ramping the rf field from $30\,$MHz down to various final evaporation frequencies. For a final evaporation frequency of $0.5\,$MHz about $2 \times 10^{5}$ $^{87}$Rb atoms are left in the Z-wire trap at a temperature of $\sim200\,$nK to produce a Bose-Einstein condensate (BEC).

The atom clouds are imaged \textit{in situ} using reflection absorption imaging~\cite{Smith2011,Surendran2015}, in which the imaging beam is sent at a small angle ($\theta \sim 2^{\circ}$) to the reflecting gold surface on the atom chip, so that two beam paths traverse the atom cloud, creating a direct image and a mirror image of the cloud (Fig.~\ref{Fig3}(a), inset). The atoms are pumped into the $|F=2, m_{F}=+2\rangle $ state and a spatially filtered $\sigma^{+}$-polarized imaging beam tuned to the $F = 2 \rightarrow F^{\prime} = 3$ cycling transition is focussed by a $50.8\,$mm-diameter achromatic lens doublet ($f_{1} = 120\,$mm, $f_{2} = 500\,$mm). The light transmitted by the atoms is imaged by the first lens which is positioned against one of the vacuum viewports at a distance $f_{1}$ from the atom cloud. The magnification is M = $f_{2}/f_{1}$, the effective pixel size in the object plane is $3.5\,\mu$m, and the measured resolution is about $9\,\mu$m. The images are recorded in a CCD camera operated in frame transfer mode.

\section{RESULTS}
\label{4}
\subsection{Bringing the Z-wire trapped atoms close to the chip surface}
\begin{figure}
\begin{centering}
\includegraphics[width=220pt]{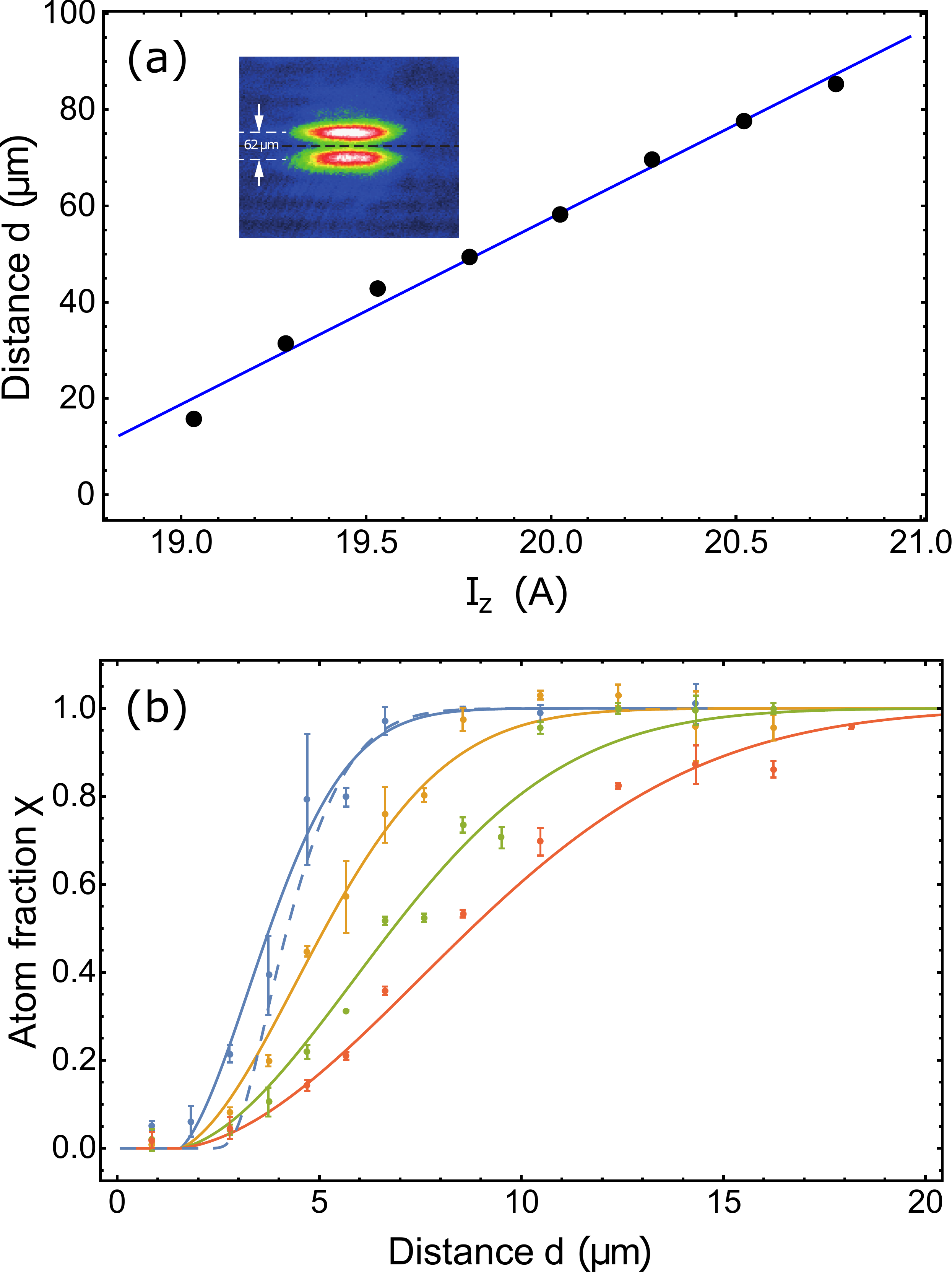}
\caption{(a) Distance calibration of the Z-wire trapped atoms close to the chip surface for $B_{x} = 52\,$G, showing  measurements of the distance $d$ from the trap centre to the gold reflecting layer on the chip surface versus Z-wire current $I_{z}$. Solid line is a linear fit: $d  = (38.8 \pm 1.6)I_{z} - (718 \pm 33)\,\mu$m, where the uncertainties are 1$\sigma$ statistical uncertainties. Inset: reflective absorption image of the atom cloud close ($31\,\mu$m) to the chip surface, showing the direct and mirror images. (b) Remaining atom fraction $\chi(d)$ versus distance $d$ of the cloud center from the chip surface for a BEC at $T\approx 200\,$nK (blue (top) points) and for a thermal cloud at $600\,$nK (orange (second) points), $1\,\mu$K (green (third) points) and $2\,\mu$K (red (bottom) points), for $B_{x} = 52\,$G. Solid curves are theoretical fits using the simple truncation model with $T\approx 190\,$nK (blue (top) line), $430\,$nK (orange (second) line), $0.85\,\mu$K (green (third) line) and $1.5\,\mu$K (red (bottom) line). The dashed blue curve for the BEC at $T\approx 200\,$nK is a theoretical fit using the 1D surface evaporation model with $T = 130\,$nK, $\tau_{el} = 0.6\,$ms.}
\par\end{centering}
\label{Fig3}
\end{figure}

To determine the distance of the centre of the Z-wire trapped atom cloud from the chip surface, we measure the separation of the centres of the direct and mirror images of clouds recorded by reflection absorption imaging (Fig.~\ref{Fig3}(a), inset). The distance between the direct and mirror images is $2d\cos \theta \approx 2d$, where $d$ is the distance of the trap centre to the chip surface. The data points (Fig.~\ref{Fig3}(a)) fit well to a straight line, where the intercept $d(I_{z}=0) = -718\,\mu$m corresponds approximately to the estimated distance of the gold mirror from the current-carrying copper wires. At very small distances from the chip surface the direct and mirror images merge into one owing to the finite size of the atom cloud and the finite resolution of the imaging system. To determine these small distances, we use an extrapolation based on the best fit to the data points in Fig.~\ref{Fig3}(a).

To investigate effects of the chip surface, we measure the fraction of remaining atoms $\chi(d)$ versus distance $d = z - 75\,$nm from the chip surface (where $z$ is the distance from the magnetic film). The atom cloud in the Z-wire trap is moved to a final position $d$, where it is held for $t_{0} = 10\,$ms, before moving back quickly to its original positon for imaging. Figure~\ref{Fig3}(b) shows the measured atom fraction $\chi(d)$ versus distance $d$ for a condensate ($T = 200\,$nK) well below the critical temperature ($T_{c} \approx 520\,$nK), and for thermal clouds at $600\,$nK, $1\,\mu$K and $2\,\mu$K. The different temperatures are obtained by changing the final evaporation frequency during the rf evaporative cooling and are measured by time of flight.

To model the atom fraction $\chi(d)$ versus distance $d$ from the chip surface, we consider the combined potential of the Z-wire magnetic trap and the attractive Casimir-Polder interaction $V(z)= V_{Z}(z)+V_{CP}(d)$, where the Z-wire trap potential is approximated by a harmonic potential $V_{z}(z)  =  1/2 M\omega_{r}^{2}(z-z_{min})^{2}$ truncated at the chip surface $z = t_{Au} + t_{SiO_{2}}$ and $V_{CP}(d)$ is given by Eq.~(\ref{CP1})~\cite{Pasquini2004}. The attractive Casimir-Polder interaction lowers the trap depth slightly to $\Delta E_{b}$ and causes the trap to disappear at a finite distance from the surface, e.g., at $d \approx 1~\mu$m for $C_{4} = 8.2 \times 10^{-56}\,$Jm$^{4}$ and $\omega_{r}/2\pi = 280\,$Hz.  The trap depth produced by the Z-wire magnetic potential plus the Casimir-Polder interaction results in a sudden truncation of the high energy tail of the Boltzmann distribution of atoms in the Z-wire trap, so that the remaining atom fraction is $\chi(d)=1-e^{-\eta}$, where $\eta = \Delta E_{b}/(k_{B}T)$ is the truncation parameter. The radial trap frequency ($\omega_{r}/2\pi = 280\,$Hz) is estimated from dipole oscillations taken over a range of distances as the Z-trap approaches the chip surface and extrapolating to the region of interest. Using $C_{4} = 8.2 \times10^{-56}\,$Jm$^{4}$, the main fitting parameter is the cloud temperature $T$, which for the four data sets in Fig.~\ref{Fig3}(b) is $190\,$nK, $430\,$nK, $0.85\,\mu$K and $1.5\,\mu$K. These values are comparable to the above temperatures measured by time of flight.

\begin{figure*}
\begin{centering}
\includegraphics[width=460pt]{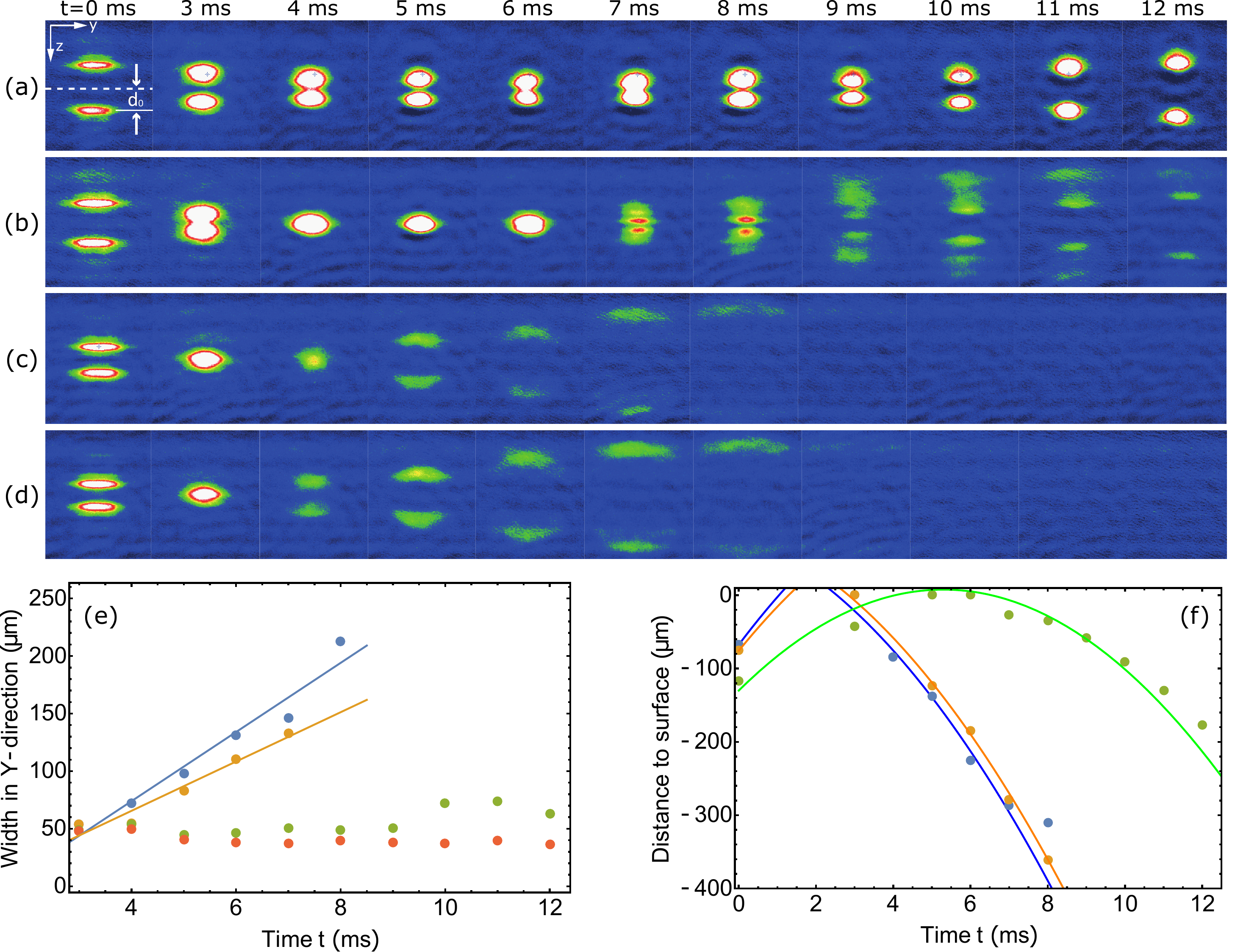}
\caption{Reflection absorption images of the time evolution of an ultracold atom cloud projected towards the magnetic lattice potential with no bias fields.  Launching position of atom cloud $d_{0}$ = (a) $145\,\mu$m, (b) $128\,\mu$m, (c) $76\,\mu$m, (d) $67\,\mu$m from the chip surface. Both direct and mirror images are visible due to the reflection absorption imaging geometry. The white dashed line in (a) indicates the position of the reflecting surface. (e),(f) Time evolution of the lateral width ($\sigma$) along $y$ and the vertical position of the ultracold atom cloud ($d$) projected towards the magnetic lattice potential. Launching positions $d_{0} = 67\,\mu$m (blue (top) points), $76\,\mu$m (orange (second) points), $128\,\mu$m (green (third) points) and $145\,\mu$m (red (bottom) points). Fitted curves in (f) (with $\sigma$ and $d$ in $\mu$m, $t$ in ms, $g = 9.8\,\mu$m/ms$^{2}$) are $d = - 67.5 + 70t - 0.5gt^{2}$ before reflection and $d = - 82.5 + 60(t + 8.1) - 0.5g(t + 8.1)^{2}$ after reflection (blue (bottom) line); $d = -75.7 + 65t - 0.5gt^{2}$ before reflection and $d = - 82.5 + 60(t + 7.8) - 0.5g(t + 7.8)^{2}$ after reflection (orange (second) line); $d = -130 + 52t - 0.5gt^{2}$ (green (top) line).}
\par\end{centering}
\label{Fig4}
\end{figure*}
The above simple truncation model can be extended to include the effect of 1D surface evaporation in which the more energetic atoms in the trap region near the chip surface preferentially escape the Z-wire trap. Using a classical 1D surface evaporation model~\cite{Lin2004}, the remaining atom fraction becomes $\chi(d)=(1-e^{-\eta})e^{-\Gamma_{ev} t_{0}}$, where $\Gamma_{ev}=f(\eta) e^{-\eta}/\tau_{el}$ is the loss rate due to 1D surface evaporation, $f(\eta) \approx 2^{-5/2} (1-\eta^{-1}+\frac{3}{2} \eta^{-2})$~\cite{Surkov1996}, $\tau_{el} =[n_{0} \sigma_{el}\overline{v}_{rel}]^{-1}$ is the elastic collision time, $\overline{v}_{rel}= \sqrt{16k_{B}T/(\pi M)}$ is the mean relative velocity, $n_{0}=\frac{N}{(2\pi)^{3/2}\sigma_{r}^{2}\sigma_{ax}}$ is the peak atom density in the Z-wire trap, $\sigma_{r,ax} = (k_{B}T/M)^{1/2}/\omega_{r,ax}$, $N$ is the number of atoms in the Z-wire trap, $\sigma_{el}=8\pi a_{s}^{2}$ is the elastic collision cross section, and $a_{s} = 5.3\,$nm is the $s$-wave scattering length for $^{87}$Rb $|F=1, m_{F}=-1\rangle$ atoms.  In Fig.~\ref{Fig3}(b) we compare fits for the 1D surface evaporation model using $T = 130\,$nK and $\tau_{el} = 0.6\,$ms (dashed blue curve) and the simple truncation model (solid blue curve) for the condensate at $200\,$nK. The discrepancy for $\chi < 0.4$ is likely due to limitations of the simple 1D surface evaporation model which ignores evaporation-induced temperature changes and the effect of collisions which can redistribute the atom directions. The redistribution of atom directions results in a larger atom loss rate and a $\chi(d)$ vs $d$ curve with a shape similar to the experimental data in Fig.~\ref{Fig3}(b)~\cite{Markle2014}.

\subsection{Interaction of ultracold atoms with the 0.7 \texorpdfstring{$\mu$m}{Lg}-period magnetic potential}
To check that the ultracold atoms can interact with the magnetic potential very close (about $100\,$nm) to the chip surface, we project an ultracold atom cloud from the Z-wire trap towards the lattice potential and monitor the reflection dynamics, similar to previous experiments with a 1D magnetic lattice potential~\cite{Singh2009,Singh2010}. This is performed for the $0.7\,\mu$m-period triangular magnetic lattice structure without bias fields, which produces a sinusoidal corrugated potential with period $\sim a$ in the $y$-direction and $\sim a/2$ in the $x$-direction~\cite{Yibo2017}.

An ultracold atom cloud at $\sim200\,$nK, i.e., below the critical temperature, is prepared in the Z-wire trap and brought to various distances $d_{0} = 145$ - $65\,\mu$m from the chip surface by ramping down $I_{z}$. The Z-wire trap is switched off suddenly by turning off $I_{z}$ and the bias field $B_{x}$. $I_{z}$ rapidly decreases to zero in $\sim 0.1\,$ms while $B_{x}$, which is produced by large Helmholtz coils, decreases slowly in $\sim10\,$ms. The resulting delay provides a momentum kick to the atom cloud, launching it vertically towards the magnetic lattice potential close to the chip surface. When the launching position is far from the chip surface, e.g., $d_{0} = 145\,\mu$m (Fig.~\ref{Fig4}(a)), the atom cloud falls down under gravity before reaching the magnetic lattice potential and no reflection is observed. Reflection signals start to appear when the launching position approaches $d_{0}=128\,\mu$m (Fig.~\ref{Fig4}(b)); both the free falling part (no lateral ($y$) expansion) and the reflected part (with lateral expansion) are observed. When $d_{0} \leq 76\,\mu$m (Fig.~\ref{Fig4}(c)-(d)), clear reflection signals are observed, which exhibit ``half-moon'' shapes due to the sinusoidal corrugation, with a lateral expansion of up to a factor of about 3.  With a 2D corrugated potential, the lateral expansion occurs in two dimensions, and since one of the directions is along the imaging beam path, the reflected cloud exhibits a half-moon shape.

Figures~\ref{Fig4}(e),(f) show the lateral width along $y$ and the vertical position of the ultracold atom cloud versus projection time $t$ for the different launching positions $d_{0}$. Without reflection, the lateral width remains almost constant at $\sim 50\,\mu$m and the trajectory of the cloud in the vertical direction fits well to a single quadratic function. For the case of reflection, the lateral width increases approximately linearly with time after reflection, with a slope corresponding to lateral velocities of 30 and $21\,\mu$m/ms for $d_{0} = 67$ and $76\,\mu$m, respectively. The fitted equations for the cloud trajectories in the caption to Fig.~\ref{Fig4} indicate (i) for $d_{0}=128\,\mu$m (green (top) curve) the atom cloud reaches its turning point after $5.3\,$ms and at about $8\,\mu$m below the chip surface, and (ii) for $d_{0}=67\,\mu$m (blue (bottom) curve) and $d_{0}=76\,\mu$m (orange (second) curve), the atom cloud interacts with the magnetic potential after $1.0\,$ms and $1.3\,$ms with an incident velocity of $60\,\mu$m/ms and $52\,\mu$m/ms and is reflected back after $2.0\,$ms and $2.6\,$ms with an exit velocity of $45$ and $45\,\mu$m/ms, respectively. When the atom cloud is launched towards a region of the magnetic film where there is no magnetic lattice structure, the atom cloud disappears almost immediately upon touching the surface.

From the above results, we conclude that the observed reflection of the atom cloud is caused by the magnetic lattice potential and that the ultracold atom cloud can interact with the short-range magnetic potential.

\subsection{Loading atoms into the 0.7 \texorpdfstring{$\mu$m}{Lg}-period triangular magnetic lattice}
The loading stage starts with a thermal cloud of $\sim 5 \times 10^{5}$ $^{87}$Rb $|F=1, m_{F}=-1\rangle$ atoms at $\sim 1\,\mu$K prepared in the Z-wire trap at $\sim 670\,\mu$m from the chip surface with $I_{z} = 38\,$A and $B_{x} = 52\,$G. Loading of the magnetic lattice is performed using a range of bias fields $B_{x} = 9$, 14, 26, 40 and $52\,$G. For $B_{x} = 52\,$G, there is no change in $B_{x}$ when the atoms are transferred from the Z-wire trap to the magnetic lattice traps and the procedure involves simply ramping down $I_{z}$. For smaller $B_{x}$, the procedure is more complex since $B_{x}$ needs to be reduced first before loading atoms into the magnetic lattice traps, which results in the Z-wire cloud being pushed away from the surface. To compensate for the change in position of the Z-wire trap, $I_{z}$ is reduced at the same time.

The atom cloud is loaded into the magnetic lattice traps by further reducing $I_{z}$ keeping $B_{x}$ fixed, which brings the atoms closer to the surface until the Z-wire trap merges smoothly with the lattice potential a few hundred nanometres from the chip surface. The ramping speed for $I_{z}$ is optimized so that it is sufficiently slow to prevent the Z-wire trapped atoms acquiring enough momentum to penetrate the magnetic lattice potential and hit the surface but not so slow that at distances very close to the chip surface the atoms are lost by surface interactions and sloshing. After the loading stage, the Z-wire cloud is brought further from the surface for imaging by rapidly ramping up $I_{z}$.
\begin{figure}[ht]
\begin{centering}
\includegraphics[width=245pt]{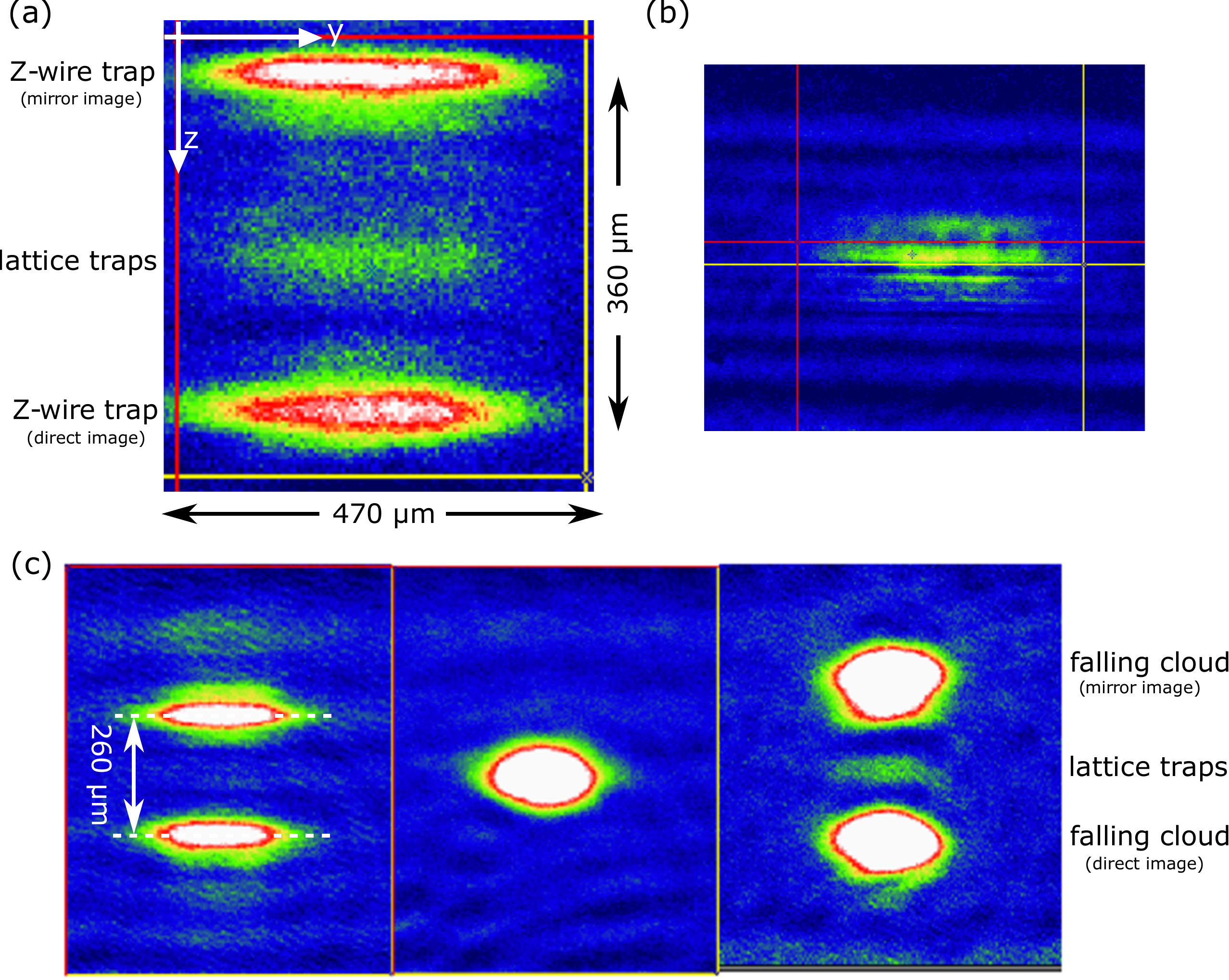}
\caption{Reflection absorption images of $^{87}$Rb $|F=1, m_{F} = -1\rangle$ atoms (a) trapped in the $0.7\,\mu$m-period triangular magnetic lattice mid-way between the direct and mirror images of the Z-wire trapped cloud, for $B_{x} = 52\,$G; (b) trapped in the $0.7\,\mu$m-period triangular magnetic lattice only, for $B_{x} = 14\,$G; and (c) after launching the atom cloud vertically towards the chip surface for times of flight of $0\,$ms (left panel), $2\,$ms (centre) and $3\,$ms (right).}
\par\end{centering}
\label{Fig5}
\end{figure}

A representative reflection absorption image is shown in Fig.~\ref{Fig5}(a) for $B_{x} = 52\,$G. The clouds at the bottom and top of the figure are the direct and mirror images of the atoms remaining in the Z-wire trap, while the smaller cloud in the middle is attributed to atoms trapped in the magnetic lattice very close to the chip surface. The direct and mirror images of the lattice trapped cloud cannot be resolved owing to their very small ($\sim 0.2\,\mu$m) separation and atoms in individual lattice sites (separated by $0.7\,\mu$m) are not resolved because of the limited resolution of the imaging system. Similar images of the small atom cloud trapped very close to the chip surface are observed for the other values of the bias field $B_{x}$.

The small atom cloud mid-way between the two larger images remains when the atoms in the Z-wire trap are removed by quickly reducing $I_{z}$ to project them vertically to hit the chip surface (Fig.~\ref{Fig5}(b)) and also when the Z-wire current is completely turned off. We estimate that typically $\sim 2 \times 10^{4}$ atoms are trapped in the magnetic lattice, initially in an area of $\sim 180\,\mu\mathrm{m}\times 13\,\mu\mathrm{m}$ (FWHM) containing about 4900 lattice sites, which corresponds to $\overline{N}_{site} \approx 4$ atoms per site. 

In a second experiment, the atom cloud is launched from a distance $d_{0} = 130\,\mu$m from the chip surface by quickly switching off both $B_{x}$ and $I_{z}$ together, so that the fast response of $I_{z}$ relative to $B_{x}$ projects the atom cloud vertically towards the magnetic lattice potential. Immediately after launching the atom cloud, small bias fields of $B_{x} = - 5.3\,$G and $B_{y} = 6\,$G are applied for 3 ms. The small negative $B_{x}$ bias field, which is produced by small fast-response Helmholtz coils, approximately cancels the residual $B_{x}$ field from the large Helmholtz coils, while the $B_{y}$ bias field creates a triangular magnetic lattice similar to the optimized lattice in Fig.~\ref{Fig1}(b). With careful optimization of the launching velocity, the atom cloud can merge with the magnetic lattice potential such that a fraction of the atoms remain trapped, while the rest fall down under gravity (Fig.~\ref{Fig5}(c), right panel). To remain trapped in the conservative potential of the magnetic lattice the atoms need to experience some dissipation which may be provided by surface evaporative cooling. After $3\,$ms time of flight the small trapped cloud appears mid-way between the direct and mirror images of the falling cloud and then disappears after a further $1.5\,$ms, which is consistent with the measured lifetime of the lattice trapped atoms (Sect.~\ref{4}D).

Further discussion about the loading of the $0.7\,\mu$m-period magnetic lattice is given in Sec.~\ref{5}.

\subsection{Lifetimes of atoms trapped in the 0.7 \texorpdfstring{$\mu$m}{Lg}-period triangular magnetic lattice}
\begin{figure}
\begin{centering}
\includegraphics[width=220pt]{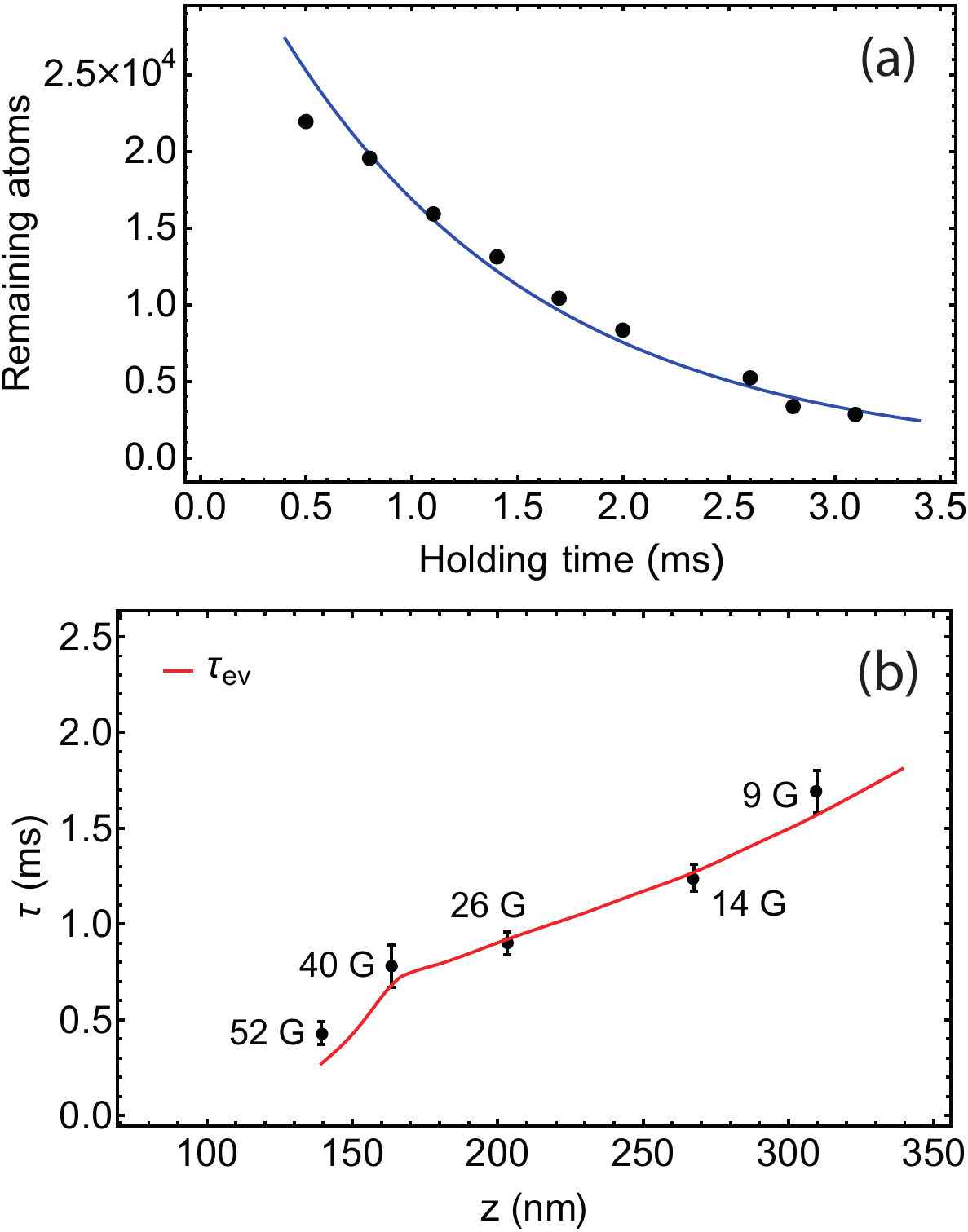}
\caption{(a) Decay curve for atoms trapped in the $0.7\,\mu$m-period triangular magnetic lattice for $B_{x} = 14\,$G. The solid line is a single exponential fit to the data corresponding to $\tau = 1.24 \pm 0.07\,$ms. Time zero is chosen arbitrarily. (b) Measured lifetimes (black points) of atoms trapped in the magnetic lattice versus distance $z$ of the lattice trap center from the magnetic film surface. The $B_{x}$ values (in G) are shown and the error bars are 1$\sigma$ statistical uncertainties. The red curve shows the calculated evaporation lifetimes $\tau_{ev}$ for $\overline{N}_{site} = 1.5$, $\eta = 4$, $\delta d = 25\,$nm and the fixed parameters given in Tables~\ref{T1} and~\ref{T2}.}
\par\end{centering}
\label{Fig6}
\end{figure}

The lifetime of the lattice trapped atoms is measured by recording the number of remaining atoms versus holding time for a range of bias fields $B_{x}$, and hence for a range of distances $z = z_{min}$ from the magnetic film surface (Table~\ref{T1}). Figure~\ref{Fig6}(a) shows a representative decay curve for $B_{x} = 14\,$G. Within our detection sensitivity, the decay curves are well fitted with a single exponential, with lifetimes varying from $0.43 \pm 0.06\,$ms for $B_{x} = 52\,$G to $1.69 \pm 0.11\,$ms for $B_{x} = 9\,$G. These lifetimes are much longer than the corresponding lattice trap periods (1 - $3\,\mu$s), and they are found to increase approximately linearly with distance $d = z - (t_{Au} + t_{SiO_{2}})$ from the chip surface over the range investigated (Fig.~\ref{Fig6}(b)). To interpret the short lifetimes and their approximately linear increase with distance $d$, we consider possible loss mechanisms.

When the thermal cloud of atoms is transferred from the Z-wire trap to the very tight magnetic lattice traps, the atoms are heated by adiabatic compression from $\sim 1\,\mu$K to an estimated initial 3 - $8\,$mK (depending on distance $d$ from the chip surface) in the magnetic lattice. Atoms with energies higher than the effective trap depth $\Delta E_{eff} = \mathrm{Min}\{\Delta E_{z}, \Delta E_{in}\}$ (Fig.~\ref{Fig1}(e)) rapidly escape the traps, resulting in a sudden truncation of the high energy tail of the Boltzmann energy distribution. We estimate that, initially, there are many ($\sim100$) atoms available for elastic collisions and evaporative cooling which provides dissipation to allow the atoms to be trapped in the conservative potential. The remaining more energetic atoms that populate the outer region of the lattice traps with energies comparable to the effective trap depth $\Delta E_{eff}$ are rapidly lost or spill over into neighboring lattice traps or are lost by rapid three-body recombination. The remaining atoms reach a quasi-equilibrium at a lower temperature $T \approx \Delta E_{eff}/(\eta k_{B})$, where $\eta$ is the truncation parameter.

\begin{figure}
\begin{centering}
\includegraphics[width=220pt]{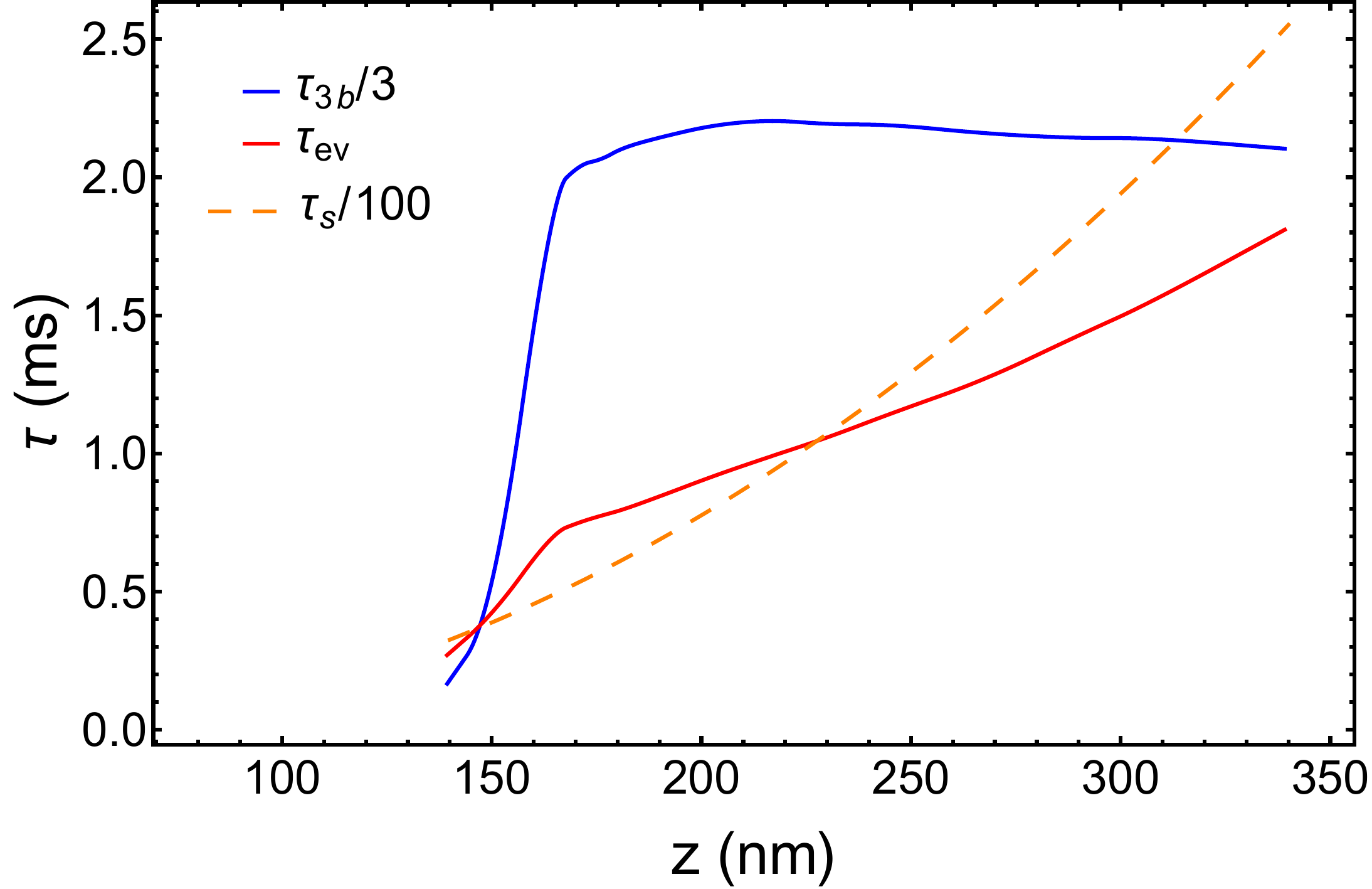}
\caption{Calculated lifetimes for evaporation $\tau_{ev}$ (red (second) curve), three-body recombination $\tau_{3b}$ (blue (top) curve) and spin flips $\tau_{s}$ (dashed orange) for $\overline{N}_{site} = 1.5$, $\eta = 4$, $\delta d = 25\,$nm and the fixed parameters given in Tables~\ref{T1} and~\ref{T2}. The curves for $\tau_{3b}$ and $\tau_{s}$ are reduced by factors of three and 100, respectively. The chip surface is located at $z = 50\,$nm.}
\par\end{centering}
\label{Fig7}
\end{figure}
In the presence of the attractive Casimir-Polder interaction, the barrier height for distances $d$ very close to the chip surface is lowest in the $z$-(vertical) direction (Table~\ref{T1}). Using the 1D evaporation model~\cite{Lin2004} in Sect.~\ref{4}A, the lifetime for one-dimensional thermal evaporation is $\tau_{ev} = \tau_{el}/[f(\eta) e^{-\eta}]$, where $\tau_{el} =[n_{0} \sigma_{el} \overline{v}_{rel}]^{-1}$, and $n_{0} = \frac{\overline{N}_{site}}{(2\pi)^{3/2}}(\frac{M}{k_{B}T})^{3/2}\overline{\omega}^{3}$ is the peak atom density in the magnetic lattice traps. According to this model, $\tau_{ev}$ scales as $\Delta E_{eff}/[\overline{\omega}^{3}\overline{N}_{site}\eta f(\eta)e^{-\eta}]$, where the truncation parameter $\eta$ is assumed to remain constant. For decreasing $B_{x} < 40\,$G (where $\Delta E_{eff} \equiv \Delta E_{z}$), the trap minima move away from the chip surface and $\overline{\omega}^{-3}$ increases at a faster rate than $\Delta E_{z}$ decreases (Table~\ref{T1}), so that $\tau_{ev}$ exhibits an almost linear increase with increasing distance $d$ from the chip surface (Fig.~\ref{Fig7}, red (second) curve). On the other hand, for increasing $B_{x} \geq 40\,$G (where $\Delta E_{eff} \equiv \Delta E_{in}$), the trap minima move very close to the chip surface and $\Delta E_{in}$ and $\overline{\omega}^{-3}$ both decrease together with decreasing $z$, resulting in a sharp decrease in $\tau_{ev}$.

A second possible loss process is three-body recombination in the very tight magnetic lattice traps. The lifetime for (non-exponential) decay by 3-body recombination is $\tau_{3b} = 1 / (K_{3}n_{0}^{2})$, where $K_{3} = 4.3(1.8) \times 10^{-29}\,$cm$^{6}$s$^{-1}$ for non-condensed $^{87}$Rb $|F = 1, m_{F} = -1\rangle$ atoms~\cite{Burt1997}. Thus, $\tau_{3b}$ scales as $\Delta E_{eff}^{3}/[\overline{\omega}^{6}\overline{N}_{site}^{2}\eta^{3}]$. For decreasing $B_{x} < 40\,$G (where $\Delta E_{eff} \equiv \Delta E_{z}$), the trap minima move away from the chip surface and $\Delta E_{z}^{3}$ decreases at about the same rate as $\overline{\omega}^{-6}$ increases (Table~\ref{T1}), so that $\tau_{3b}$ remains almost constant for distances  $z  > 170\,$nm (Fig.~\ref{Fig7}, blue (top) curve). For increasing $B_{x} \geq 40\,$G (where $\Delta E_{eff} \equiv \Delta E_{in}$), the trap minima move very close to the chip surface and $\Delta E_{in}^{3}$ and $\overline{\omega}^{-6}$ both decrease strongly together with decreasing $z$, resulting in a rapid decrease in $\tau_{3b}$.

A further possible loss process can result from spin flips caused by Johnson magnetic noise from the gold conducting layer on the magnetic film~\cite{Lin2004,Treutlein2008,Jones2003}. The spin-flip lifetime is given by $\tau_{s} = \frac{256\pi \hbar^{2}d}{3\mu_{0}^{2}\mu_{B}^{2}\sigma k_{B}T g(d,t_{Au},\delta)}$for state $|F=1,m_{F}=-1\rangle$~\cite{Treutlein2008}, where $g(d,t_{Au},\delta) \approx  t_{Au}/(t_{Au} + d)$ for $\delta \gg$ Max$\{ d,t_{Au} \}$~\cite{Henkel2005}; $\delta = \sqrt{2/(\sigma \mu_{0}\omega_{L})}$ is the skin depth at the spin-flip transition frequency $\omega_{L} = m_{F}g_{F}\mu_{B}B_{IP}/\hbar$; $\sigma$ is the electrical conductivity of the conducting layer; and $\mu_{0}$ is the vacuum permeability. For $t_{Au} = 50\,$nm, we obtain spin-flip lifetimes (Fig.~\ref{Fig7}, dashed orange curve) that are much longer than the measured trap lifetimes, for example, $\tau_{s} = 48\,$ms and $230\,$ms for $d = 110\,$nm and $290\,$nm, respectively.

The calculated one-dimensional evaporation lifetime $\tau_{ev}$ versus distance (Fig.~\ref{Fig7}, red (second) curve) has a positive slope, given approximately by $\Delta E_{eff}/(\overline{\omega}^{3}d)$, which closely matches the slope of the measured lifetime versus distance (Fig.~\ref{Fig6}), with no adjustable parameters. On the other hand, the calculated $\tau_{3b}$ versus distance (Fig.~\ref{Fig7}, blue (top) curve) remains almost constant for $z > 170\,$nm. This suggests that the dominant loss mechanism limiting the trap lifetimes is one-dimensional thermal evaporation, rather than three-body recombination or spin flips due to Johnson magnetic noise. With thermal evaporation, one might expect some atoms to remain in the lattice traps for times much longer than $1\,$ms. Within our detection sensitivity, there is no indication of a non-exponential tail in the decay curves, e.g., Fig.~\ref{Fig6}(a).

The red curve in Fig.~\ref{Fig6}(b) shows the calculated evaporation lifetime $\tau_{ev}$ with fitted scaling parameters $\overline{N}_{site} = 1.5$, $\eta = 4$, a fitted offset $\delta d = 25\,$nm (see below) and the fixed parameters given in Tables~\ref{T1} and~\ref{T2}. To obtain a reasonable fit such that the evaporation lifetime is much shorter than the three-body recombination lifetime requires a value $\overline{N}_{site} \approx 1.5$ which is smaller than the $\overline{N}_{site} \approx 4$ estimated from the number of atoms ($\sim 2 \times 10^{4}$) initially trapped in $\sim 4900$ lattice sites. The smaller value of $\overline{N}_{site} \approx 1.5$ could be a result of atoms spilling over into neighboring lattice sites during the initial transfer of atoms from the Z-wire trap into the tight magnetic lattice traps, so that more than 4900 lattice sites are occupied at the time of measurement of the atom number and/or it could be a result of uncertainties in the size of the Z-trap cloud or the total number of lattice trapped atoms. An average of $1.5$ atoms per lattice site over the occupied lattice is consistent with the end-product of rapid three-body recombination prior to the observation period, leaving zero, one or two atoms on any given site.

To obtain a reasonable fit to the measured lifetimes at very small distances $d$ from the chip surface, where the calculated lifetime is very sensitive to the distance $d$ due to the Casimir-Polder interaction, requires either the calculated $C_{4} = 8.2 \times 10^{-56}\,$Jm$^{4}$ to be smaller by an order of magnitude or the calculated distances of the trapped atoms from the chip surface $d = z_{min}-(t_{Au} + t_{SiO_{2}})$ to be larger by $\delta d \approx 25$ nm. The above $C_{4}$ value is expected to be accurate to within $\sim 40\%$ based on the level of agreement between the calculated $C_{4}$ value and the measured value~\cite{Stehle2011} for a dielectric sapphire surface film. A value of $\delta d = 25\,$nm is within the estimated uncertainty $\left(^{+40}_{-30}\,\mathrm{nm}\right)$ in $d = z_{min}-( t_{Au} + t_{SiO_{2}})$ for $B_{x} = 40\,$G and $52\,$G, which has contributions from a systematic error of about $+$$10\,$nm due to the effect of the $20\,$nm-deep etching of the magnetic film and estimated uncertainties in $t_{Au} + t_{SiO_{2}}$ ($\pm 5\,$nm) and $z_{min}$ ($\pm 25\,$nm) and the effect of the estimated uncertainty in $C_{4}$ ($\pm 2\,$nm).

\section{DISCUSSION}
\label{5}
The measured lifetimes of the atoms trapped in the $0.7\,\mu$m-period magnetic lattice are short, 0.4 - $1.7\,$ms for distances $d = 90$ - $260\,$nm from the chip surface, and need to be increased to enable quantum tunneling. For example, for an atomic system trapped in a $0.7\,\mu$m-period square lattice with a trap depth of $12E_{r}\sim20\,$mG (where $E_{r}=\hbar^{2}k^{2}/(2M)$ is the recoil energy), the Bose-Hubbard model for the Mott-insulator transition predicts that at the critical point, which occurs at $(J/U)_{c} \sim 0.06$~\cite{Batrouni2002}, the tunneling matrix element $J/k_{B} = 0.82\,$nK, the on-site interaction energy $U/k_{B} = 14\,$nK, and the tunneling time is $9\,$ms~\cite{Bloch2008,Bakr2009}.

Our model calculations suggest that the short lifetimes of the atoms trapped in the magnetic lattice are currently limited mainly by losses due to one-dimensional thermal evaporation following transfer of the thermal atom cloud from the Z-wire trap into the very tight magnetic lattice traps, rather than by fundamental loss processes such as surface interactions, three-body recombination or spin flips due to Johnson magnetic noise. Therefore, it should be technically feasible to reach longer lifetimes in the magnetic lattice traps by reducing the effect of one-dimensional thermal evaporation following the loading process. One possible way is to increase the distance of the trapped atoms from the magnetic surface, for example, by using a thicker magnetic film and/or by using an optimized triangular magnetic lattice with $z_{min}\approx a \approx 700\,$nm. However, increasing $z_{min}$ reduces not only the mean trap frequency but also the trap depth, thereby resulting in only a marginal increase in the trap lifetime, as exemplified in Fig~\ref{Fig6}(b).

A bigger gain is likely to come from improving the transfer of atoms from the Z-wire trap to the very tight magnetic lattice traps. Heating due to adiabatic compression during transfer of the thermal cloud to the magnetic lattice traps could be reduced by loading the atoms from a magnetic trap with trap frequency higher than $\sim100\,$Hz. Trap frequencies as high as $5\,$kHz~\cite{Lin2004} or even tens of kilohertz~\cite{Jacqmin2012} have previously been achieved for a current-carrying conductor microtrap on an atom chip. A further gain in transfer efficiency could be obtained by ensuring that the direction of the trap bottom field ($B_{IP}$) of the magnetic lattice traps is aligned with that of the Z-wire trap. It should also be possible to reduce heating due to adiabatic compression if a BEC, rather than a thermal cloud, can be loaded directly from the Z-wire trap into the magnetic lattice.

If trap lifetimes $\sim 100\,$ms can be achieved, losses due to spin flips caused by Johnson magnetic noise may become significant (Fig.~\ref{Fig7}, dashed orange curve). Such losses could be reduced, for example, by replacing the reflecting $50\,$nm gold layer ($\rho = 0.22 \times 10^{-7}\,\Omega$m) on the chip with a reflecting material with higher resistivity such as palladium ($\rho = 1.05 \times 10^{-7}\,\Omega$m) and by operating at larger distances from the conducting layer, as discussed above.

To gain a more complete understanding of the loss processes presently limiting the trap lifetimes, it would be informative to study magnetic lattices with periods in between $0.7\,\mu$m and $10\,\mu$m (for which trap lifetimes of $10\,$s have been achieved~\cite{Surendran2015}).

\section{SUMMARY AND CONCLUSIONS}
\label{6}
We have demonstrated trapping of ultracold $^{87}$Rb atoms in a $0.7\,\mu$m-period triangular magnetic lattice on an atom chip based on the following observations:

(\romannum{1})  The atom cloud is found to interact with the magnetic lattice potential very close to the chip surface when it is projected vertically towards the surface.

(\romannum{2}) A small atom cloud appears mid-way between the direct and mirror images of the Z-trapped atom cloud when it is brought very close to the chip surface. The small cloud remains when the atoms remaining in the Z-wire trap are removed and when the Z-wire current is completely turned off.

(\romannum{3}) A small atom cloud also appears very close to the chip surface when a cloud of atoms is projected vertically from the Z-wire trap with optimized velocity to almost touch the chip surface.

(\romannum{4}) The lifetimes of the small atom cloud (0.4 - $1.7\,$ms) are much longer than the corresponding lattice trap periods (1 - $3\,\mu$s) and increase significantly with increasing distance from the chip surface, approximately in accordance with model calculations.

Our model calculations suggest that the trap lifetimes are currently limited mainly by losses due to one-dimensional thermal evaporation following transfer of atoms from the Z-wire trap to the very tight magnetic lattice traps, rather than by fundamental loss processes such as surface interactions, three-body recombination or spin flips due to Johnson magnetic noise. It should be feasible to overcome one-dimensional thermal evaporation losses by improving the transfer of atoms from the Z-wire trap to the very tight magnetic lattice traps, for example, by loading the atoms from a magnetic trap with higher trap frequency.

The trapping of atoms in a $0.7\,\mu$m-period magnetic lattice represents a significant step towards using magnetic lattices for quantum tunneling experiments and to simulate condensed matter and many-body phenomena in nontrivial lattice geometries. To the best of our knowledge, the trapping of atoms at distances of about $100\,$nm from the chip surface and at trap frequencies as high as $800\,$kHz represents new territory for trapping ultracold atoms.

\section*{ACKNOWLEDGMENTS}
We are indebted to Shannon Whitlock, Russell McLean, Saulius Juodkazis and Peter Kr\"{u}ger for fruitful discussions. We thank Pierette Michaux for fabricating early versions of the magnetic lattice structures and James Wang for assistance with the magnetic force/atomic force microscope measurements. The electron beam lithography was performed at the Melbourne Centre for Nanofabrication (MCN) in the Victorian Node of the Australian National Fabrication Facility (ANFF). The atom chip was fabricated using the nanofabrication facility at Swinburne University. Funding from the Australian Research Council (Discovery Project Grant No. DP130101160) is acknowledged.


\begin{thebibliography}{10}

\bibitem{Yibo2016}Y. Wang, P. Surendran, S. Jose, T. Tran, I. Herrera, S. Whitlock, R. McLean, A. Sidorov, and P. Hannaford, Sci. Bulletin \textbf{61}, 1097 (2016).

\bibitem{Schmied2010}R. Schmied, D. Leibfried, R. J. C. Spreeuw, and S. Whitlock, New J. Phys. \textbf{12}, 103029 	(2010).

\bibitem{Singh2008}M. Singh, M. Volk, A. Akulshin, A. Sidorov, R. McLean, and P. Hannaford, J. Phys. B \textbf{41}, 065301 (2008).

\bibitem{Jose2014}S. Jose, P. Surendran, Y. Wang, I. Herrera, L. Krzemien, S. Whitlock, R. McLean, A. Sidorov, and P. Hannaford, Phys. Rev. A \textbf{89}, 051602(R) (2014).

\bibitem{Surendran2015}P. Surendran, S. Jose, Y. Wang, I. Herrera, H. Hu, X. Liu, S. Whitlock, R. McLean, A. Sidorov, and P. Hannaford, Phys. Rev. A \textbf{91}, 023605 (2015).

\bibitem{Gerritsma2007}R. Gerritsma, S. Whitlock, T. Fernholz, H. Schlatter, J. A. Luigjes, J. -U. Thiele, J. B. Goedkoop, and R. J. C. Spreeuw, Phys. Rev. A \textbf{76}, 033408 (2007).

\bibitem{Whitlock2009}S. Whitlock, R. Gerritsma, T. Fernholz, and R. J. C. Spreeuw, New J. Phys. \textbf{11}, 023021 (2009).

\bibitem{Leung2011}V. Y. F. Leung, A. Tauschinsky, N. J. van Druten, and R. J. C. Spreeuw, Quant. Inf. Process. \textbf{10}, 955 (2011).

\bibitem{Herrera2016}I. Herrera, Y. Wang, P. Michaux, D. Nissen, P. Surendran, S. Juodkazis, S. Whitlock, R. McLean, A. Sidorov, M. Albrecht, and P. Hannaford,
J. Phys. D \textbf{48}, 115002 (2015).

\bibitem{Leung2014}V. Y. F. Leung, D. R. M. Pijn, H. Schlatter, L. Torralbo-Campo, A. L. La Rooij, G. B. Mulder, J. Naber, M. L. Soudijn, A. Tauschinsky, C. Abarbanel, B. Hadad, E. Golan, R. Folman, and R. J. C. Spreeuw, Rev. Sci. Instrum. \textbf{85}, 053102 (2014).

\bibitem{Bloch2008}I. Bloch, J. Dalibard, and W. Zwerger, Rev. Mod. Phys. \textbf{80}, 885 (2008).

\bibitem{Bakr2009}W. S. Bakr, J. I. Gillen, A. Peng, S. F\"{o}lling, and M. Greiner, Nature (London) \textbf{462}, 74 (2009).

\bibitem{Pasquini2004}T. A. Pasquini, Y. Shin, C. Sanner, M. Saba, A. Schirotzek, D. E. Pritchard, and W. Ketterle, Phys. Rev. Lett. \textbf{93}, 223201 (2004).

\bibitem{Lin2004}Y. Lin. I. Teper, C. Chin, and V. Vuletic, Phys. Rev. Lett. \textbf{92}, 050404 (2004).

\bibitem{Yan1997}Z. Yan, A. Dalgarno, and J. F. Babb, Phys. Rev. A \textbf{55}, 2882 (1997).

\bibitem{Stark2015}M. St\"{a}rk, F. Schlickeiser, D. Nissen, B. Hebler, P. Graus, D. Hinzke, E. Scheer, P. Leiderer, M. Fonin, M. Albrecht, U. Nowak, and J. Boneberg,
 Nanotechnology \textbf{26}, 205302 (2015).

\bibitem{Stinson1990}D. G. Stinson and S.-C. Shin, J. Appl. Phys. \textbf{67}, 4459 (1990).

\bibitem{Yibo2017}Y. Wang, PhD thesis, Swinburne University of Technology (2017).

\bibitem{Squires2011}M. B. Squires, J. A. Stickney, E. J. Carlson, P. M. Baker, W. R. Buchwald, S. Wentzell, and S. M. Miller, Rev. Sci. Instr. \textbf{82}, 023101 (2011).

\bibitem{Burt1997}E. A. Burt, R. W. Ghrist, C. J. Myatt, M. J. Holland, E. A. Cornell, and C. E. Wieman, Phys. Rev. Lett. \textbf{79}, 337 (1997).

\bibitem{Soding1999}J. S\"{o}ding, D. Gu\'{e}ry-Odelin, P. Desbiolles, F. Chevy, H. Inamori, and J. Dalibard, Appl. Phys. B \textbf{69}, 257 (1999).

\bibitem{Smith2011}D. A. Smith, S. Aigner, S. Hofferberth, M. Gring, M. Andersson, S. Wildermuth, P. Kr\"{u}ger, S. Schneider, T. Schumm, and J. Schmiedmayer,
Opt. Express \textbf{19}, 8471 (2011).

\bibitem{Surkov1996}E. L. Surkov, J. T. M. Walraven, and G. V. Shlyapnikov, Phys. Rev. A \textbf{53}, 3403 (1996).

\bibitem{Markle2014}J. M\"{a}rkle, A. J. Allen, P. Federsel, B. Jetter, A. G\"{u}nther, J. Fort\'{a}gh, N. P. Proukakis, and T. E. Judd, Phys. Rev. A \textbf{90}, 023614 (2014).

\bibitem{Singh2009}M. Singh, R. McLean, A. Sidorov, and P. Hannaford, Phys. Rev. A \textbf{79}, 053407 (2009).

\bibitem{Singh2010}M. Singh and P. Hannaford, Phys. Rev. A \textbf{82}, 013416 (2010).

\bibitem{Treutlein2008}P. Treutlein, PhD Thesis, Ludwig-Maximilians University Munich (2008).

\bibitem{Jones2003}M. P. A. Jones, C. J. Vale, D. Sahagun, B. V. Hall, and E. A. Hinds, Phys. Rev. Lett. \textbf{91}, 080401 (2003).

\bibitem{Henkel2005}C. Henkel, Eur. J. Phys. D. \textbf{35}, 59 (2005).

\bibitem{Stehle2011}C. Stehle, H. Bender, C. Zimmermann, D. Kern, M. Fleischer, and S. Slama, Nat. Photon. \textbf{5}, 494 (2011).

\bibitem{Jacqmin2012}T. Jacqmin, B. Fang, T. Berrada, T. Roscilde, and I. Bouchoule, Phys. Rev. A. \textbf{86}, 043626 (2012).

\bibitem{Batrouni2002}G. Batrouni, V. Rousseau, R. Scalettar, M. Rigol, A. Muramatsu, P. Denteneer, and M. Troyer, Phys. Rev. Lett. \textbf{89}, 117203 (2002).

\end{thebibliography}
\end{document}